\documentstyle[11pt,twoside,jltp,epsf]{article}

\title{Superconductivity and Spin Fluctuations}

\author{D.J.~Scalapino\address{
Department of Physics,\\
University of California,\\
Santa Barbara, CA 93106-9530 USA\\
email: djs@spock.physics.ucsb.edu}}

\runninghead{D.J.~Scalapino}{Superconductivity and Spin Fluctuations}
       
\begin{document}

\begin{abstract}
The organizers of the Memorial Session for Herman Rietschel asked that
I review some of the history of the interplay of superconductivity and
spin fluctuations. Initially, Berk and Schrieffer showed how
paramagnon spin fluctuations could suppress superconductivity in
nearly-ferromagnetic materials. Following this, Rietschel and various
co-workers wrote a number of papers in which they investigated the
role of spin fluctuations in reducing the $T_c$ of various
electron-phonon superconductors. Paramagnon spin fluctuations are also
believed to provide the $p$-wave pairing mechanism responsible for the
superfluid phases of $^3He$. More recently, antiferromagnetic spin
fluctuations have been proposed as the mechanism for $d$-wave pairing
in the heavy-fermion superconductors and in some organic materials as
well as possibly the high-$T_c$ cuprates.  Here I will review some of
this early history and discuss some of the things we have learned more
recently from numerical simulations.

PACS numbers: 74
\end{abstract}

\maketitle

\vspace{0.3in}

The interplay of magnetic spin fluctuations and superconductivity has
an interesting history, beginning as a way to understand the
suppression of $T_c$ in some of the traditional metals, then providing
a mechanism for $p$-wave pairing in $^3He$, and finally as a suggested
$d$-wave pairing mechanism for some organic superconductors,
heavy-fermion systems, and possibly the high-$T_c$ cuprates. Here I
will review some of the early history and discuss what has been
learned in recent years from numerical studies.

\begin{figure}[ht]
\centerline{\epsfysize=1.7in \epsfbox{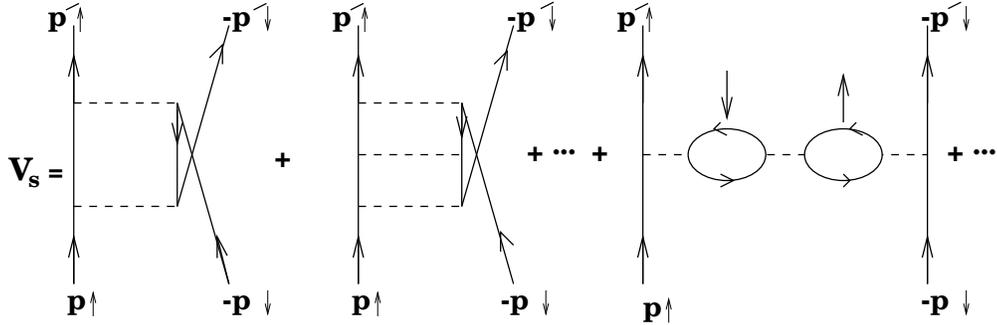}}
\vskip .10 in
\caption{Diagrams representing the Berk-Schrieffer [1]
spin-fluctuation mediated pairing interaction in the singlet channel.}
\end{figure}

In 1966 Berk and Schrieffer \cite{BS66} and Doniach and Engelsberg
\cite{DE66} discussed the behavior of nearly-ferromagnetic systems in
terms of paramagnon spin fluctuations.  Berk and Schrieffer showed how
paramagnon fluctuations could lead to a strong suppression of the
superconducting transition temperature in materials such as
$Pd$. Figure~1 illustrates the Berk-Schrieffer paramagnon mediated
interaction in the singlet pairing channel for a Hubbard model with an
on-site renormalized interaction $\bar U n_{i\uparrow}
n_{i\downarrow}$. The particle-hole graphs on the left contain the
contribution from the transverse spin-fluctuations and the bubble
graphs to the right contain the longitudinal spin fluctuations. For a
nearly-ferromagnetic system, this singlet channel interaction has the
form
\begin{equation}
V_s(q,\omega) \cong \frac{3}{2}\ \frac{\bar U^2 \chi_0 
(q,\omega)}{1-\bar U\chi_0(q,\omega)}
\label{one}
\end{equation}
with
\begin{equation}
\chi_0(q,\omega) = \int \frac{d^3p}{(2\pi)^3}
\ \frac{f(\varepsilon_{p+q})-f(\varepsilon_p)}{\omega -(\varepsilon_{p+q}
-\varepsilon_p)+i\delta}
\label{two}
\end{equation}
the usual Lindhard function.
A schematic plot of $V_s(q,0)$ is shown in Fig.~2. The peak at $q=0$ is set
by the Stoner enhancement factor $(1-\bar U\chi_0(0))^{-1}$. 

\begin{figure}[ht]
\centerline{\epsfysize=2.2 in \epsfbox{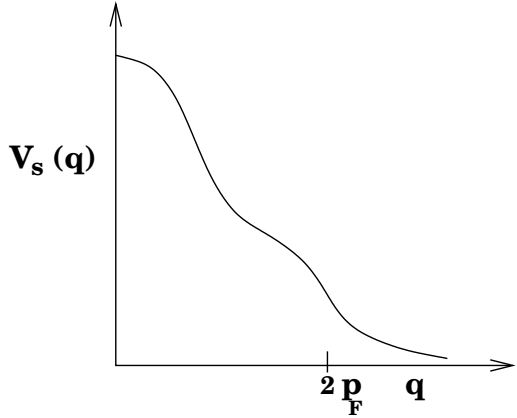}}
\vskip .10 in
\caption{Schematic plot of $V_s(q)$ versus $q$ for a spherical fermi
surface.}
\end{figure}

Now, just as the usual electron-phonon interaction is characterized by a
dimensionless parameter $\lambda$, the paramagnon spin-fluctuation
interaction has
\begin{equation}
\lambda_{SF} = - \int^\infty_0 \frac{\left\langle
ImV_s(q,w)\right\rangle}{w}\ dw = - Re \left\langle V_s(q,0)\right\rangle.
\label{three}
\end{equation}
Here the bracket indicates an average of the momentum transfer over
the fermi surface appropriate for a constant s-wave gap. From the
sketch of $V_s(q,0)$ in Fig.~2, one clearly sees that $\lambda_{SF}$
is negative and the paramagnon spin-fluctuation exchange suppresses
$s$-wave superconductivity.

Thus, as Berk and Schrieffer \cite{BS66} showed, one could understand
why nearly-ferromagnetic metals such as $Pd$ were not
superconducting. In the late '70's and early '80's, Herman Rietschel
and his co-workers \cite{RW79RWR80R81} used this approach to discuss
the possibility that metals which had significantly smaller Stoner
enhancement factors might nevertheless have their $T_c$ values limited
by paramagnon exchange. For example, they noted that band structure
and frozen phonon calculations of the electron-phonon coupling
suggested that $Nb$ and $V$ might be expected to have transition
temperatures as high as 18K. They then discussed the possibility that
while paramagnon exchange did not suppress the $T_c$'s of $Nb$ and $V$
to zero, it could provide an explanation for why they were well below
18K. In particular, they concluded that since a ``high electronic
density of states $N(0)$ favors the occurrence of spin fluctuations,
paramagnon effects are one, if not the limiting factor for high
superconducting transition temperatures.''  Rietschel
et.~al.~\cite{GRJPM85} extended these ideas to discuss a variety of
materials including $VN$, $Nb_{1-x}V_xN$ and $V_2Zr$. To appreciate
this work, one needs to realize that this occurred at a time when the
success of the Eliashberg theory \cite{Eli60} suggested that ab-initio
calculations of $T_c$ were possible and furthermore that such
calculations might provide an understanding of the maximum
superconducting $T_c$ that could be reached. In particular, while
increasing the electron density of states $N(0)$ might increase the
effective electron-phonon coupling parameter, it could also increase
the suppression arising from the Stoner-enhanced spin fluctuations and
thus there could be some optimum situation and hence maximum $T_c$.

\begin{figure}[ht]
\centerline{\epsfysize=2.1 in \epsfbox{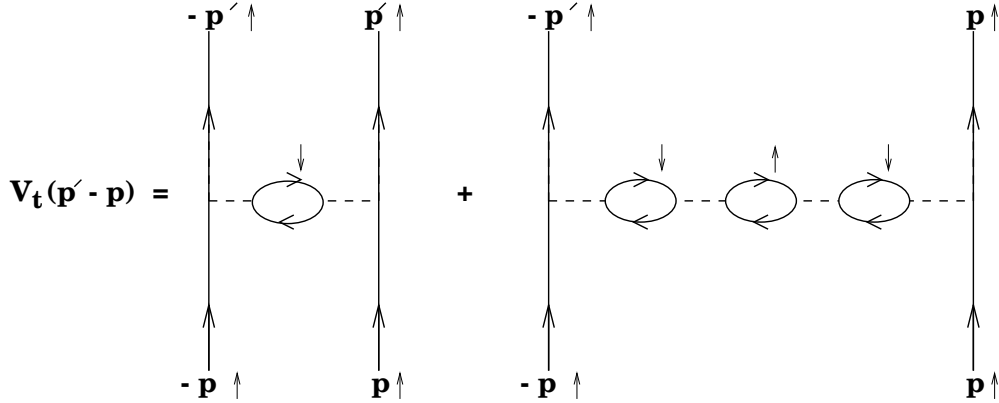}}
\vskip .10 in
\caption{Diagrams representing the paramagnon spin-fluctuation exchange
interaction in the $S_Z=1$ triplet-pairing channel.}
\end{figure}

\begin{figure}[ht]
\centerline{\epsfysize=2.2 in \epsfbox{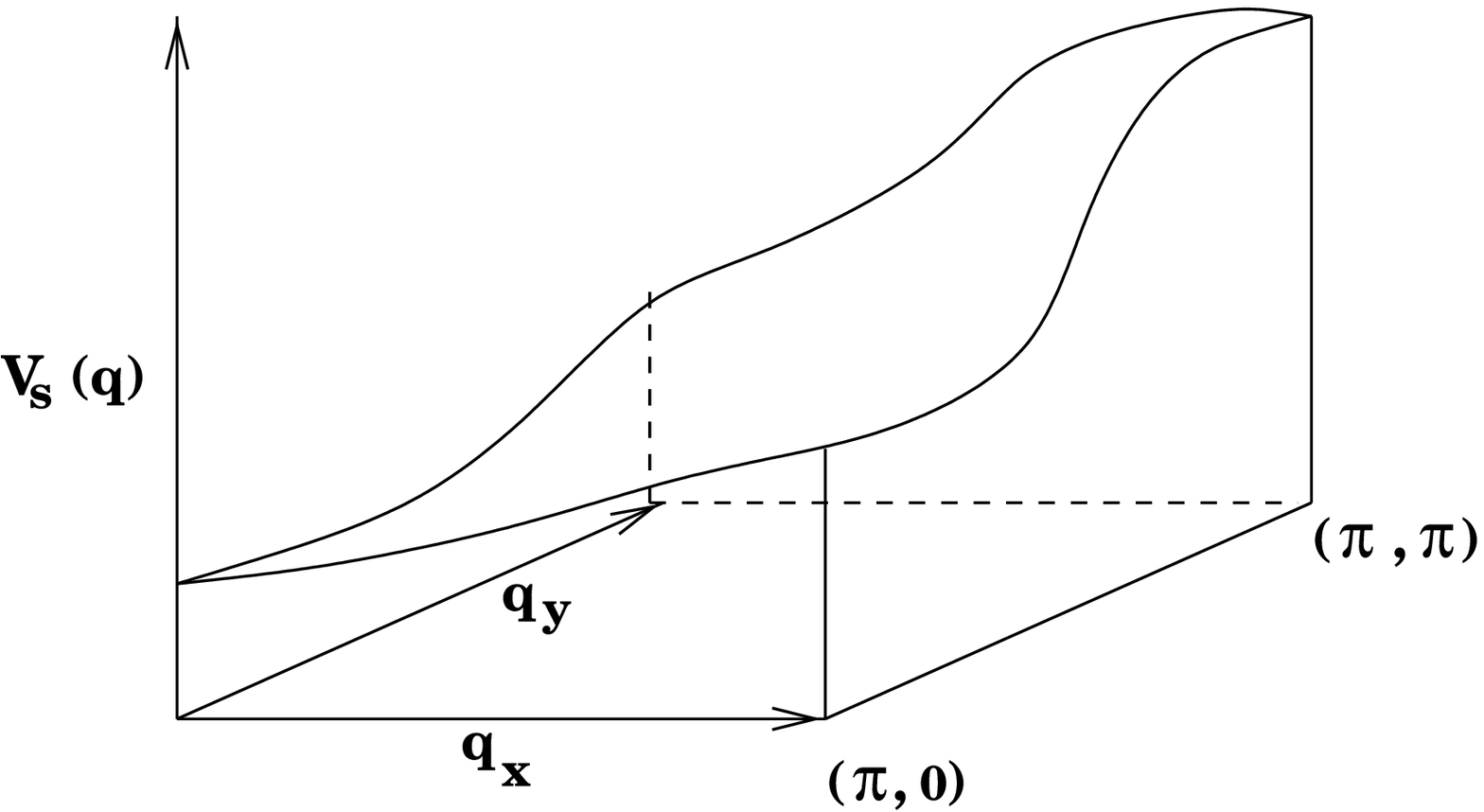}}
\vskip .10in
\caption{Sketch of $V_s(q)$ versus $q$ for a two-dimensional system with
short-range antiferromagnetic spin fluctuations.}
\end{figure}

Now, previous to this, in 1971, Layzer and Fay \cite{LF71} had
suggested that paramagnetic spin fluctuations could provide a $p$-wave
pairing mechanism for $^3He$, explicitly implementing an earlier
suggestion for odd-channel pairing by Emery \cite{Eme64}. The
paramagnon exchange interaction in the triplet channel, illustrated
for $S_z=1$ in Fig.~3, gives an effective pairing interaction
\begin{equation}
V_t \cong -\frac{\bar U^2}{2}\ \frac{\chi_0(q,\omega)}{1-\bar
U\chi_0(q,\omega)}.
\label{four}
\end{equation}
In this approximation, $V_t(q)$ is just {\it minus} one-third
$V_s(q)$.  Averaging this interaction over the fermi surface with a
p-wave form factor for the gap gives a positive (attractive) effective
pairing interaction strength because $V_s(q)$ is peaked at small
momentum transfers.
%so that averaging it over the
%fermi surface gives a positive (attractive) effective pairing
%interaction strength.
After the discovery of superfluid $^3He$ in 1973, Anderson and
Brinkman \cite{AB73} showed how the feedback of the superconducting
state modified the paramagnetic spin fluctuations favoring the
formation of the anisotropic $p$-wave Anderson-Morel \cite{AM61} state
over the Balian-Werthamer state \cite{BW63} when the spin fluctuations
are strong.

\begin{figure}[ht]
\centerline{\epsfysize=2.63 in \epsfbox{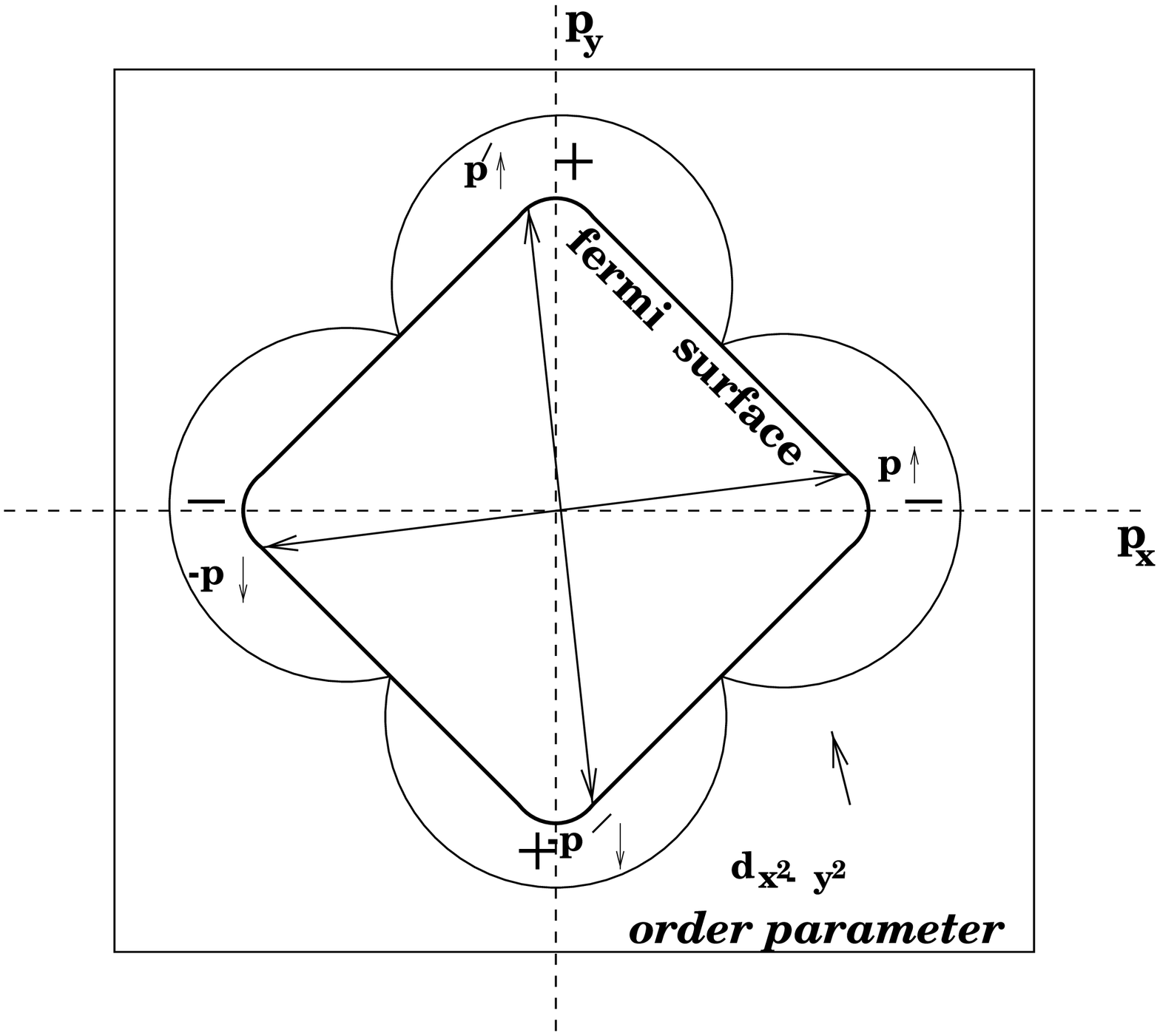}}
\vskip .10 in
\caption{Illustration showing how a $d$-wave gap can provide a
solution of the BCS gap eq.~(\ref{five}) for a pairing
interaction which increases at large momentum transfer like the type
illustrated in Fig.~4.}
\end{figure}

Finally, moving forward to 1986, a paper by Emery appeared in which he
suggested that back-scattering from spin fluctuations could lead to
the pairing of holes on neighboring organic stacks in the Beckgaard
salts \cite{Eme86}. In addition, three papers were submitted in June
of '86 which argued that antiferromagnetic spin fluctuations might
mediate $d$-wave pairing in the heavy-fermion materials
\cite{Cyr86,SLH86,MS-RV86}. To see how this can happen, Fig.~4 shows
the singlet Berk-Schrieffer interaction, eq.~(\ref{one}), versus $q$
for a two-dimensional system with antiferromagnetic fluctuations.  In
this case $V_s(q)$ peaks near $(\pi, \pi)$ rather than at
$q=(0,0)$. This may come about from band-structure nesting effects as
it does in the weak-coupling treatment of the nearly half-filled
Hubbard model \cite{SLH86,BSW89} or because of strong-coupling,
short-range valence bond correlations such as in the 2-leg Hubbard or
$t$-$J$ ladders \cite{DRS92,RTT95,NSW}. Basically, it is the
short-range spin fluctuations that drive the pairing.

There are several ways to understand how such an interaction can give
rise to pairing. First consider the usual BCS equation
\begin{equation}
\Delta_p = - \sum_{p^\prime}
\ \frac{V (p-p^\prime) \Delta_{p^\prime}}{2E_{p^\prime}}.
\label{five}
\end{equation}

\begin{figure}[ht]
\centerline{\epsfysize=2.2 in \epsfbox{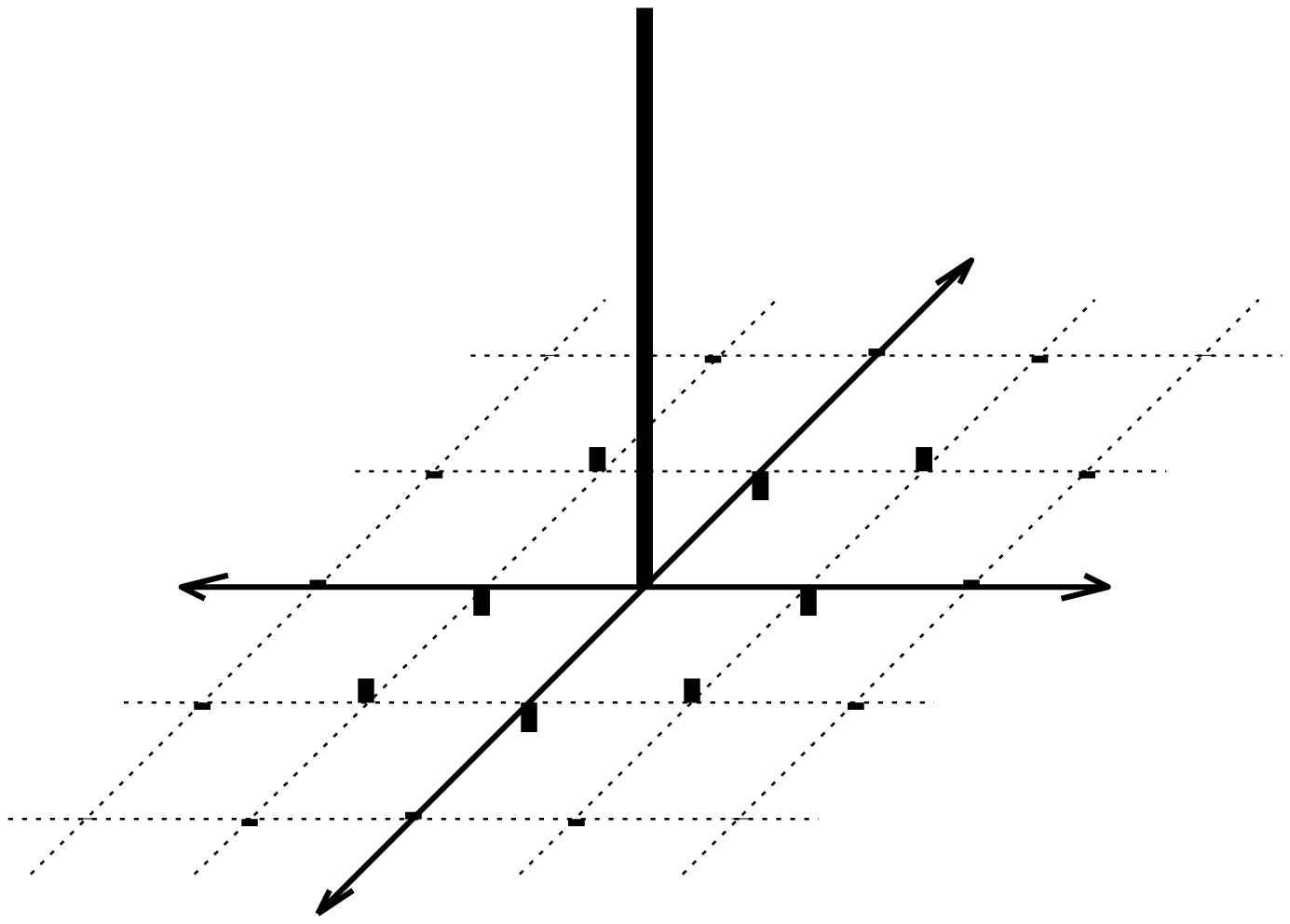}}
\vskip .10 in
\caption{Fourier transform of the singlet-pairing interaction
$V_s(q)$ of Fig.~4 arising from the short-range antiferromagnetic spin
fluctuations on a square lattice.  Here one member of the singlet pair
is located at the origin and the other at a surrounding site
$\ell$. The potential is strongly repulsive for both electrons on the
same site, as shown by the large positive bar at the origin.  However,
the potential is attractive on near-neighbor sites.}
\end{figure}

\noindent
For the traditional electron-phonon superconductors, $V$ is negative
and slowly varying over the fermi surface. This leads to a gap,
$\Delta_p$, which has the same sign over the fermi surface and which
is only weakly anisotropic. However, for an interaction which becomes
more positive at large momentum transfer, and a large fermi surface,
the gap must change sign on the fermi surface as schematically
illustrated in Fig.~5 in order to satisfy the BCS gap,
eq.~(\ref{five}). That is, suppose $\Delta_{p^\prime}$ is positive for
$p^\prime$ near $(0,\pi)$. Then the strong scattering for
${\mathbf{p}-\mathbf{p}^\prime} \simeq (\pi,\pi)$ produces a negative
gap $\Delta_p$ at $\mathbf{p} = (\pi,0)$ according to eq.~(\ref{five})
so that a $d_{x^2-y^2}$ gap of the form $\Delta_p=\Delta_0 (\cos\, p_x
+ \cos\, p_y)$ provides a solution of the BCS gap equation.

Alternatively, a spatial Fourier transform of the interaction
$V_s(q,\omega=0)$
\begin{equation}
V_s(\ell) = \sum_q e^{i\vec q \cdot \vec \ell}\, V_s(q,\omega=0)
\label{six}
\end{equation}
gives the result sketched in Fig.~6. Here $\vec \ell$ is the
separation of the electrons making up the pair.  The pairing
interaction $V_s(\ell)$ is clearly repulsive on site but becomes
attractive on near-neighbor as well as some longer-range sites. The
electrons making up the pair can bind when they spatially arrange
themselves to take advantage of the attractive regions.  Note that the
electrons making up the pair come from states within the
spin-fluctuation energy of the fermi surface so that a
$d_{x^2-y^2}$-wave is formed (for the 2D case) rather than an extended
$s$-wave.

This was the kind of picture that was evolving in mid 1986 and the
hope was that further measurements would tell whether
superconductivity in the heavy-fermion materials and possibly some of
the organic materials were indeed mediated by a magnetic
spin-fluctuation mechanism.  However, there was already at that time a
publication, submitted in April 17 of that year by Bednorz and Muller
\cite{BM86}, which would change the direction of research and lead to
an intense questioning of the validity of the antiferromagnetic
spin-fluctuation pairing mechanism. In fact, to this day, there is no
widely agreed-upon pairing mechanism for the high-$T_c$ cuprate
problem although there is renewed interest in the antiferromagnetic
spin-fluctuation exchange as the mechanism for pairing in the
heavy-fermion materials as well as the organics.

\begin{figure}[ht]
%\centerline{\epsfysize=2.5 in \epsfbox{fig7a.eps}}
%\centerline{\epsfysize=2.5 in \epsfbox{fig7b.eps}}
\centerline{\epsfysize=6.0cm \epsffile[-30 184 544 598] {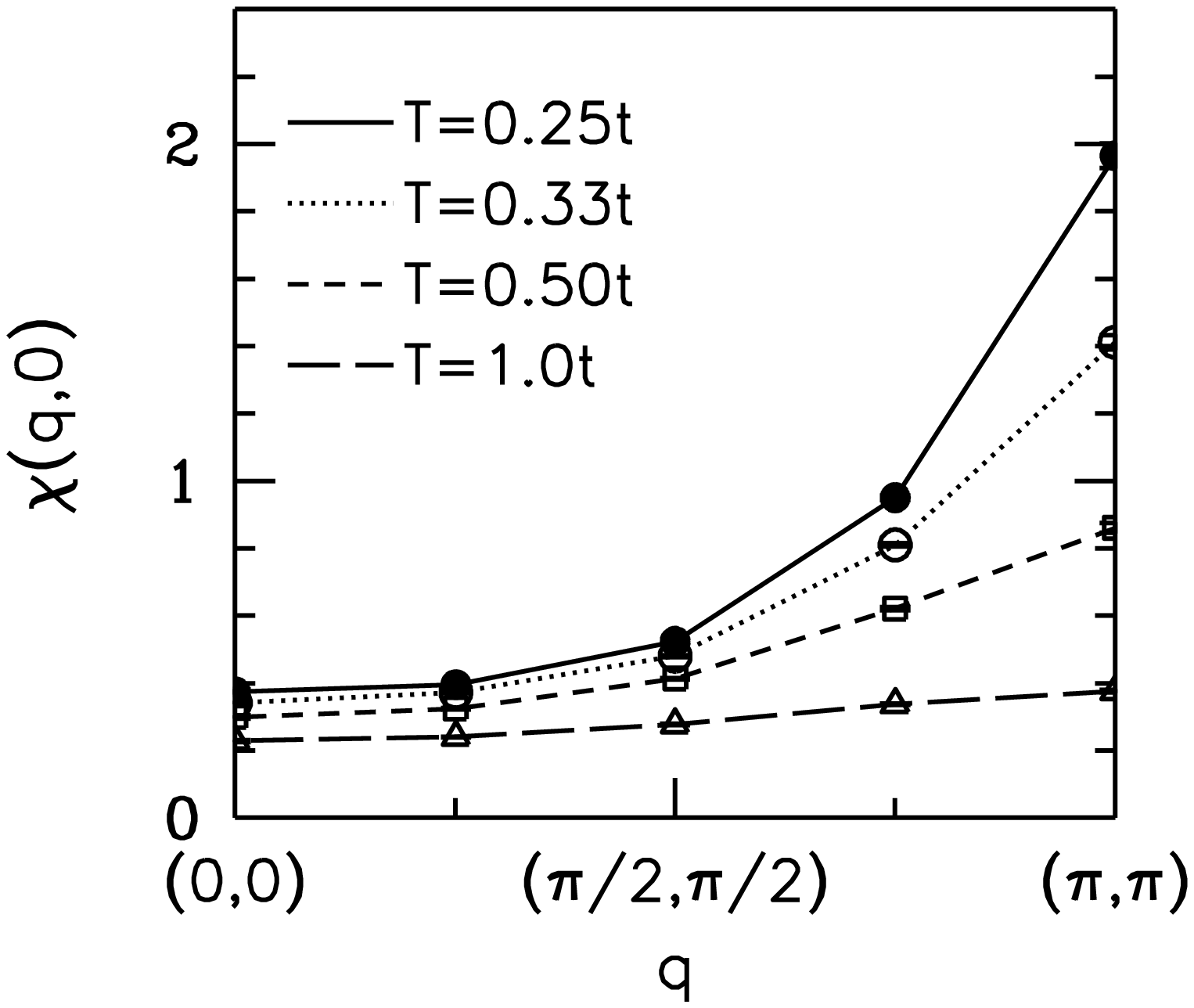}
\epsfysize=6.0cm \epsffile[98 184 672 598] {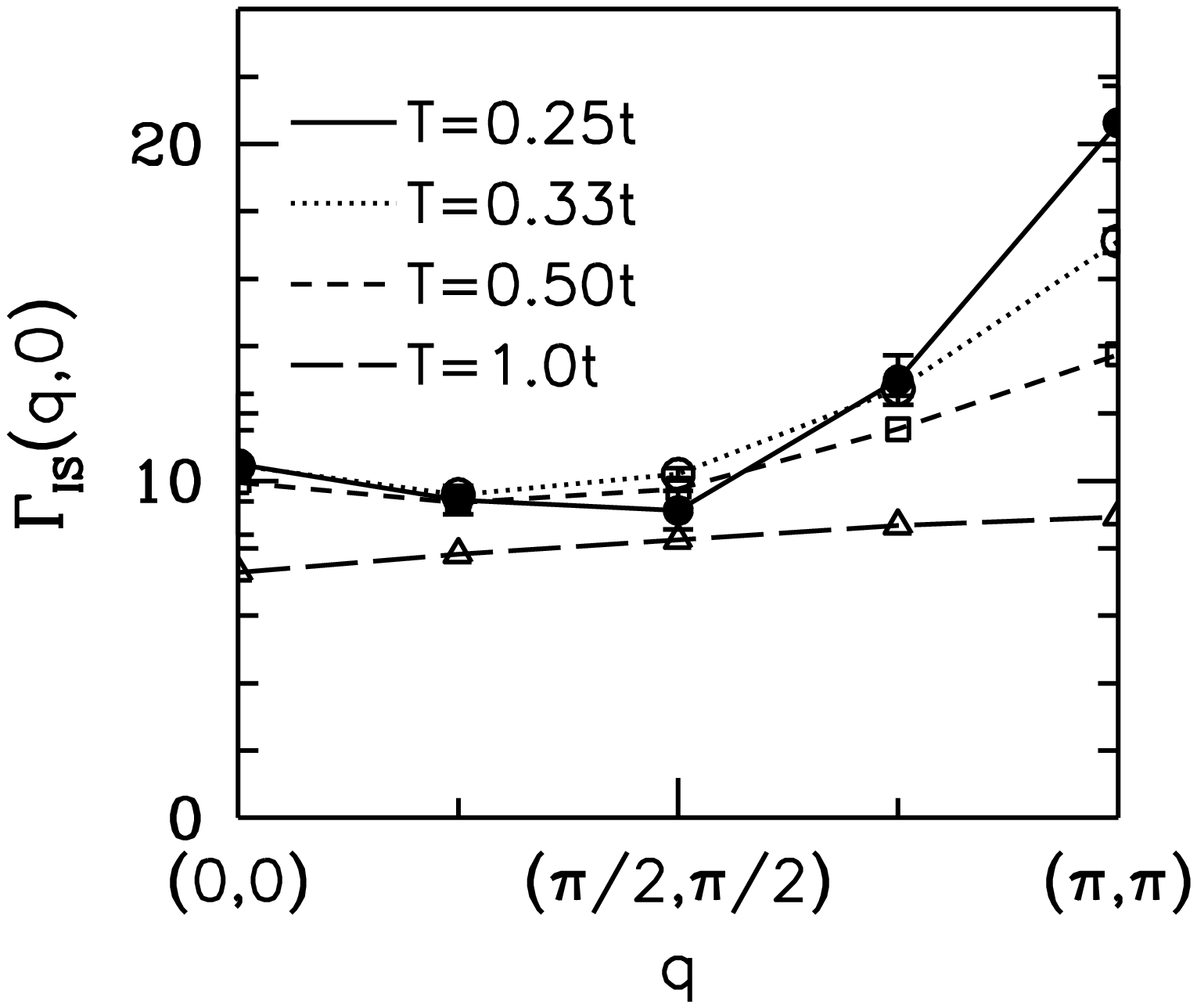}}
\vskip .26in
\caption{a) Magnetic susceptibility $\chi({\mathbf q})$ versus
{\bf q} along the (1,1) direction for $\omega_m=0$ and various
temperatures.  These results are for an $8\times8$ lattice with
$U/t=4$ and a filling $\langle n\rangle=0.875$.  b) The effective
pairing interaction in the singlet channel versus $q$ along the
$(1,1)$ direction for the same parameters as (a).  From ref.~[26].  }
\end{figure}

In addition, with the passage of time, we have, in fact, learned more
about the structure of the pairing mechanism for various basic systems
such as the Hubbard and $t$-$J$ models. There have been Gutzwiller
variational calculations \cite{GJR87,YS88}, auxiliary boson mean-field
treatments \cite{IDHR88,KL88}, conserving fluctuation exchange
diagramatic calculations \cite{BSW89,MTU90}, as well as various
phenomenological spin-fluctuation mediated pairing calculations
\cite{MBP91}.  Here I will focus on some of what we have learned about
the pairing interaction from numerical calculations. Monte Carlo
calculations \cite{BSW93} for both the doped 2D Hubbard model and the
2-leg Hubbard ladder \cite{DS97} find that the effective
particle-particle interaction in the singlet channel peaks at large
momentum transfer.  Figure~7 shows Monte Carlo results for a doped
$8\times 8$ Hubbard lattice with $U=8t$ and a site filling $\langle
n\rangle = .875$. The peak in the interaction at large momentum
transfers Fig.~7b evolves as the temperature $T$ is lowered in a
similar manner to the evolution of the peak in the magnetic spin
susceptibility $\chi(q)$, Fig.~7a. The structure of the pairing
interaction in Fig.~7b, shown for momentum transfer along the $(1,1)$
direction, is similar to the increase at large momentum transfer found
in weak-coupling calculations, Fig.~4.

A similar data set for the two-leg Hubbard ladder \cite{DS97} is shown
in Fig.~8. In this case one is dealing with a spin-gapped system in
which the antiferromagnetic correlations decay exponentially and the
spin-fluctuations can be pictured as local fluctuations of the rung
and leg singlet bonds. Nevertheless, as seen in Fig.~8 the temperature
dependence of the large-momentum structure of the pairing interaction
is similar to the spin susceptibility and the overall structure is
similar to the results for the 2D Hubbard model.

\begin{figure}[ht]
%\centerline{\epsfysize=2.5 in \epsfbox{fig8a.eps}}
%\centerline{\epsfysize=2.5 in \epsfbox{fig8b.eps}}
\centerline{\epsfysize=6.0cm \epsffile[-30 184 544 598] {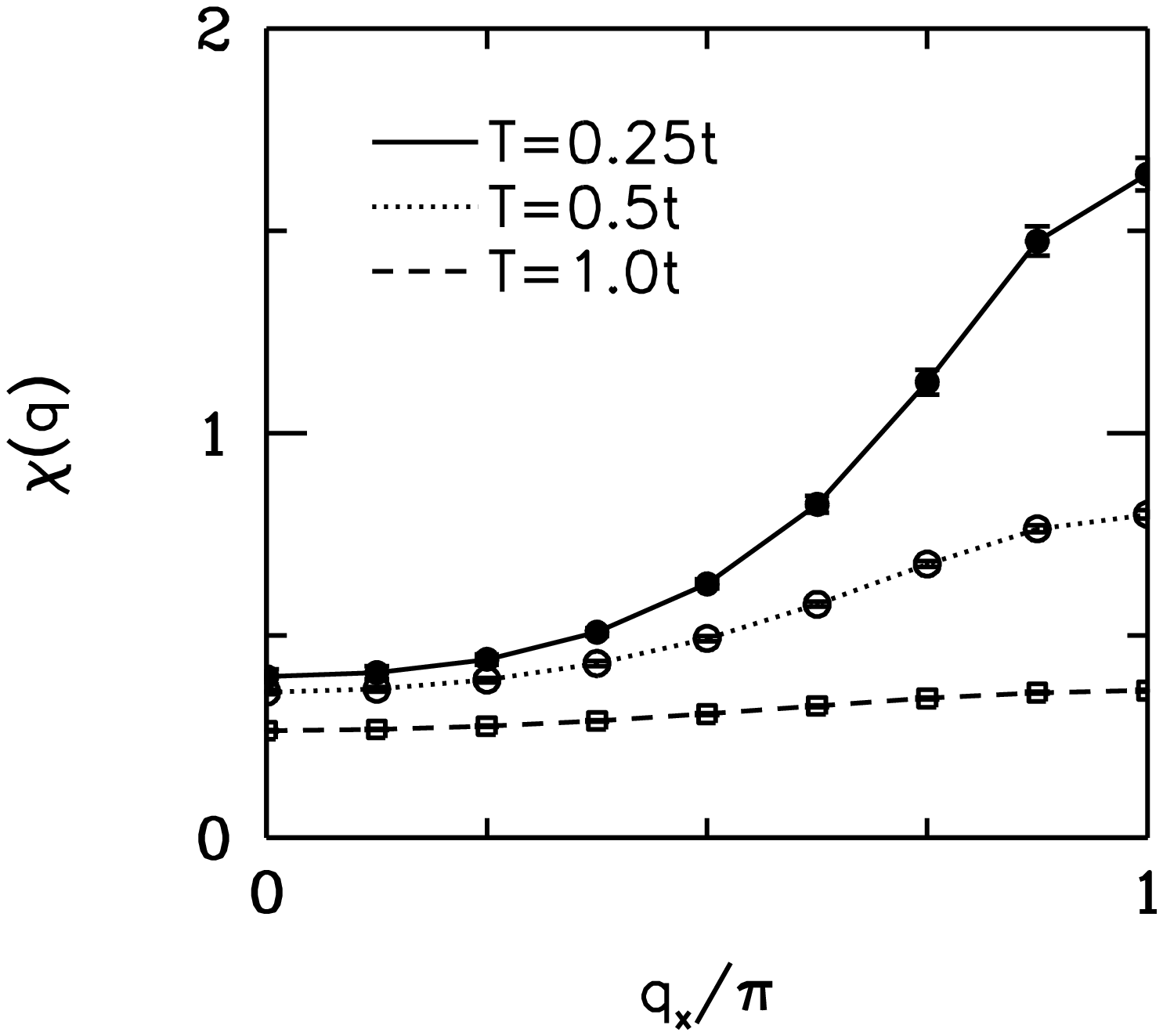}
\epsfysize=6.0cm \epsffile[98 184 672 598] {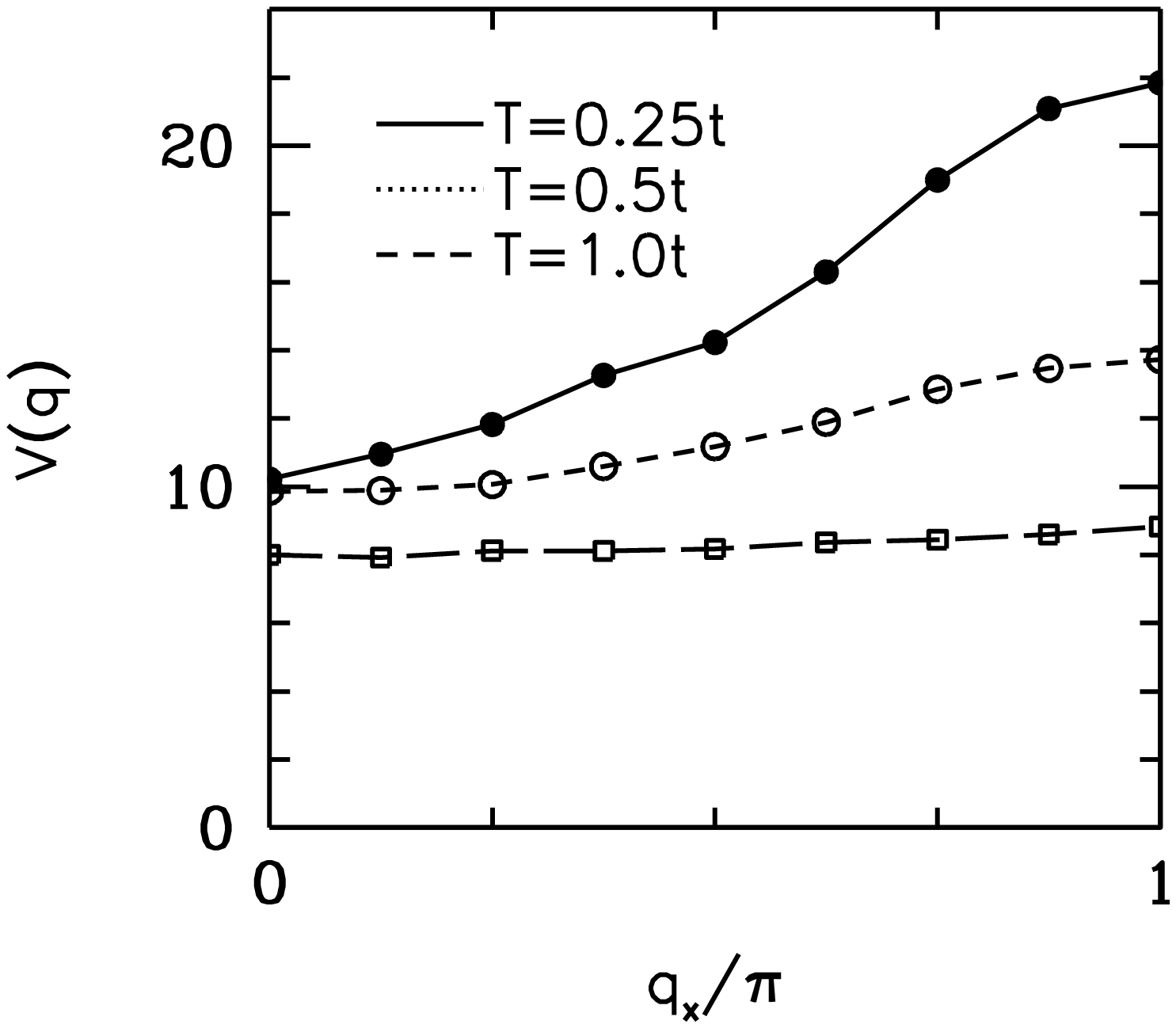}}
\vskip .26 in
\caption{a) Momentum dependence of the magnetic susceptibility
$\chi(\mathbf{q})$ for a 2-leg ladder with $U=4t, \langle n\rangle =
0.875$ and $t_\perp = 1.5t$.  Here $q_y=\pi$ and $\chi(\mathbf{q})$ is
plotted as a function of $q_x$. b) Momentum dependence of the
effective interaction $V(\mathbf{q})$ for $U=4t, \langle n\rangle =
0.875$ and $t_\perp =1.5t$. Here $V(\mathbf{q})$ is measured in units
of $t, q_y=\pi$ and $V(\mathbf{q})$ is plotted as a function of $q_x$.
From ref.~[27].  }
\end{figure}

Using these effective two-particle interactions along with the Monte
Carlo calculation of the single-particle Green's function, we have
solved the particle-particle Bethe-Saltpeter equations \cite{BSW94}
and shown that the leading eigenvalue is associated with the
$d_{x^2-y^2}$-wave (or in the case of the 2-leg ladder, the
$d_{x^2-y^2}$-wave-like) pairing instability. However, the
temperatures for which we can carry out these Monte Carlo calculations
are limited by the fermion sign problem and at the lowest temperatures
we have reached, approximately half the effective exchange energy, the
$d$-wave eigenvalue is only of order 0.3. A pairing instability would
occur if this eigenvalue reaches 1, where there would be a
Kosterlitz-Thouless transition for the 2D system.  Other types of
Monte Carlo calculations \cite{GOG99} suggested that the ground state
of the basic 2D Hubbard model with only a near-neighbor hopping $t$
and an onsite Coulomb interaction $U$ has only short-range, $d$-wave
correlations. Numerical density matrix renormalization group
calculations do find power law $d_{x^2-y^2}$-like pairing correlations
for the doped 2-leg Hubbard ladder, but in order for these
correlations to be dominant over the charge-density correlations one
needs to have a ratio of the rung-to-leg hopping larger than unity
\cite{NBSZ97} or consider other modifications of the basic model
involving next-near-neighbor hopping or an additional contribution to
the exchange interaction. Of course, both of these will be present in
the actual system.  In particular, in the $CuO$ system, an additional
exchange coupling, beyond that which is present in a one-band Hubbard
model arises from processes in which an intermediate state has two
holes on an $O$. We have found, in fact, that the addition of such an
exchange term to the Hubbard ladder gives rise to a significant
enhancement of the pairing correlations \cite{DSW99}.

Thus, although there remain strong theoretical divisions \cite{And97}
regarding the role of magnetic spin fluctuations as a $d$-wave pairing
mechanism, our numerical results support this
possibility. Furthermore, recent experimental results on the
heavy-fermion materials \cite{Mat98,JHA99} $CePd_2Si_2$, $CeIn_3$, and
$UPd_2A\ell_3$ and the layered organic superconductors \cite{McK97}
$\kappa$-$(BEDT$-$TTF)_2X$ provide experimental support for this type
of pairing mechanism in these systems.  The ultimate resolution of the
high-$T_c$ cuprate problem remains to be sorted out. In this
connection, it is encouraging that numerical calculations find
evidence for both $d_{x^2-y^2}$ pair formation and domain wall
formation which appear to arise from the interplay of short-range,
spin-correlations and the kinetic energy of the doped holes in the
$t$-$J$ model.

\section*{ACKNOWLEDGMENTS}

The work reported here was carried out with N.~Bulut and S.R.~White.  I
would like to acknowledge support from the DOE under grant \# DOE85-ER45197.

%\end{references}
\end{document}